# Tuning Yttrium Σ7(0001) Twist Grain Boundary Properties through Segregation and Co−segregation of Low Neutron Absorption Elements: First-Principles Insights

Guanlin Lyu [a], Yuguo Sun [a], Panpan Gao [a], Ping Qian [a, *]

[a] Department of Physics, University of Science and Technology Beijing, Beijing, 100083, China

[*] Corresponding author

**Corresponding author:   Ping Qian**

Address:    30 Xueyuan Road, Haidian District, Beijing China

E−mail:   qianping@ustb.edu.cn

**Other Authors:**

Guanlin Lyu,    Email: guanlinlyu@outlook.com

Yuguo Sun,    Email: guoguo@xs.ustb.edu.cn

Panpan Gao,    Email: gaopanpan@ustb.edu.cn

**Abstract**

Elements with low thermal neutron absorption cross−sections are ideal for enhancing structural materials in nuclear systems. In this study, We systematically investigate the segregation and co−segregation behaviors of eleven elements at the Σ7(0001) twist grain boundary in yttrium and their effects on stability and strength. The Σ7(0001) grain boundary exhibits weakening, with fracture occurring preferentially along path I. Segregation energy calculations show that Si, Cu, Cr, Mo and Fe prefer interstitial sites, while others occupy substitutional ones. Si, Al, Zn, Cu, Mg and Fe stabilize the boundary, while Mo, Fe, Si, Cr, Cu, Nb and Ti strengthen it, with Si offering the most balanced improvement. Co−segregation studies reveal that Si induces the enrichment of other solutes at the boundary, promoting synergistic stabilization and turning embrittling elements (Al, Mg, Zn, Zr) into strengthening agents. Electronic structure analysis shows that Si–Y covalent bonds enhance electron localization, and Si+Mg co–segregation optimizes electronic distribution through metallic–covalent cooperation, significantly improving fracture resistance. The density of states analysis indicates new low–energy deep states in the Si, and Si+Al, Si+Mg systems, which lower grain boundary energy and improve stability. This study provides guidance for designing high–performance, low−neutron−absorption Y−based alloys.

## 1. Introduction

With the rapid development of advanced fields such as aerospace, nuclear energy, and new energy technologies, structural materials are increasingly required to operate under extreme service environments involving severe stress, irradiation, and corrosion. These conditions pose significant challenges to their mechanical performance and structural stability [1,2]. In polycrystalline metals, grain boundaries (GBs), as one of the most common and energetically unfavorable structural defects, exhibit atomic arrangements that differ markedly from those in the bulk lattice. Consequently, GBs often serve as preferred sites for plastic deformation, crack initiation, and corrosion damage, thereby playing a critical role in determining the overall service performance of materials [3–10]. Therefore, achieving a fundamental understanding of GB behavior and effectively tuning their structure and properties have become central scientific issues in the development of next−generation high−performance structural materials.

In recent years, grain boundary segregation engineering (GBSE) has emerged as an effective interface design strategy, offering new pathways for developing alloys with simultaneously high stability and strength [11–14]. Studies have shown that the segregation behavior of specific solute elements at GBs can significantly modify the local chemical environment and bonding characteristics near the boundary, thus influencing interfacial cohesion and mechanical response [15–25]. First−principles calculations based on density functional theory provide a powerful approach for understanding these phenomena, as they can reveal, at the electronic structure level, the atomic reconstruction, bonding evolution, and intrinsic effects of solute−induced segregation on GB stability and fracture behavior [26–31].

Extensive first–principles studies have systematically investigated the segregation effects and strengthening mechanisms of various solute elements at representative metallic grain boundaries. For instance, Mg segregation enhances the stability of Al Σ5(310) and Σ9(221) boundaries but reduces their cohesion, whereas Cu segregation significantly strengthens the boundaries through the formation of Cu–Al covalent bonds, while Na segregation induces pronounced embrittlement [16,17]. Segregation of Zr, Nb, and Mo is found to substantially strengthen the Cu Σ5(310) boundary, primarily due to the strong interaction between their d−electrons and those of Cu boundary [18]. In the vicinity of the Cu Σ11 [110]($1\bar{1}3$) grain boundary, elements such as In, Ca, Zn, Ag, Zr, and Sn exhibit significant segregation tendencies; however, only Zr enhances the boundary strength. Moreover, Zr can induce the enrichment of typically reluctant elements, such as Cr and Co, further improving both the cohesion and stability of the boundary [21]. In the Fe Σ5(310) grain boundary, the segregation of elements such as Ni, Co, Ti, V, Mn, Nb, Cr, Mo, W, and Re exhibits a pronounced strengthening effect, which is primarily governed by the combined influence of atomic size and electronic structure [23]. Similarly, in the Ni Σ5(012) symmetrical tilt grain boundary, segregation of V increases the maximum tensile strength by approximately 17% [26]. At the Y {10$\bar{1}$1} twin boundary, segregation of Ru and Pt enhances the boundary strength, while Ru additionally shows a tendency to induce Bi segregation [30]. Xue et al. systematically investigated the segregation behavior of 30 solute elements at the Zr {10$\bar{1}$1} twin boundary and found that V, Cr, Nb, Mo, Ta, and W act as effective strengthening agents, whereas Ag, Bi, Au, and Sn lead to grain boundary embrittlement [31]. These studies demonstrate that first−principles calculations offer unique advantages—such as element−level controllability, well−defined configurations, and interpretable results—making them a powerful tool for uncovering grain boundary segregation mechanisms, predicting strengthening trends, and guiding experimental design and alloy composition optimization [16,21].

Yttrium (Y) metal, characterized by its hexagonal close–packed crystal structure, high–temperature stability, and excellent irradiation resistance, has been recognized as a promising structural material candidate for nuclear energy and high–technology applications [32–34]. In the design of Y–based alloys, the addition of alloying elements must satisfy the critical constraint of maintaining a low thermal neutron absorption cross−section, which makes it particularly important to investigate elements with favorable neutron–physical properties. However, systematic studies on the segregation and co–segregation behaviors of such elements at Y grain boundaries, as well as their effects on grain boundary strength and electronic structure modulation, remain scarce.

Based on this, the present study employs the Σ7(0001) twist grain boundary with relatively low boundary energy in hcp–Y metal as a model system. Excluding noble and highly toxic elements (such as Au, Be, and Cd), a series of candidate alloying elements with potential applications in nuclear engineering and low thermal neutron absorption cross sections—Al, Cr, Cu, Fe, Mg, Mo, Nb, Si, Ti, Zn, and Zr—were selected. Using first−principles calculations based on density functional theory, we systematically investigated their segregation and co−segregation behaviors at this grain boundary. By evaluating the effects of solute segregation on grain boundary cohesive strength, electronic structure, and bonding characteristics, the underlying microscopic mechanisms governing the modulation of interfacial properties were revealed. This study aims to provide an in−depth understanding, at both the atomic and electronic levels, of how solute elements regulate the grain boundary behavior in Y−based alloys, thereby offering theoretical guidance for interfacial optimization and compositional design of Y alloys under extreme service conditions.

## 2. Models and Methods

### 2.1 First−principles Calculations

The present study is based on first−principles calculations within the framework of density functional theory, as implemented in the Vienna Ab initio Simulation Package [35,36], which is widely used in condensed matter physics and materials simulations. The construction and visualization of atomic models were performed using the VESTA and Atomic Simulation Environment tools [37,38]. The interactions between ions and electrons were described using the projector augmented−wave method [39,40], and the pseudopotentials were adopted from the standard VASP library. The electronic exchange–correlation energy was treated using the Perdew–Burke–Ernzerhof functional within the framework of the generalized gradient approximation [41,42], which has been extensively validated for accurately describing the structural and energetic properties of metals and their alloys.

After careful convergence testing, Brillouin zone sampling was performed using a Monkhorst–Pack k−point mesh centered at the Γ point, with a grid density corresponding to KSPACING = 0.2 Å$^{-1}$ [43]. The plane−wave energy cutoff was set to 400 eV. During electronic self−consistency, the energy convergence threshold was set to $1\times10^{-6}$ eV, and ionic relaxations were conducted using the conjugate−gradient algorithm. The structures were considered fully relaxed when the residual forces on all atoms were below 0.01 eV·Å$^{-1}$. For systems containing magnetic elements, spin polarization was explicitly taken into account.

**2.2 Grain boundary and Surface model**

The grain boundary (GB) model was constructed based on the hcp–Y Σ7(0001) twist grain boundary structure proposed by Zheng et al. [7]. After full structural relaxation, the calculated GB energy was approximately 0.30 J·m$^{-2}$. To balance computational accuracy and efficiency, two atomic layers farthest from the boundary plane were removed following the strategy adopted by Christensen et al. for the hcp–Zr Σ7(0001) GB [44], without altering the structural characteristics of the interface. The resulting GB energy remained 0.30 J·m$^{-2}$. Although this value is slightly higher than the reported 0.22 J·m$^{-2}$ [7], the discrepancy likely arises from differences in VASP versions and pseudopotentials. Nevertheless, the relatively low energy indicates good thermodynamic stability of the constructed model, validating its reliability as the basis for subsequent segregation studies.

The hcp–Y Σ7(0001) grain boundary (GB) is formed by rotating two crystal grains by 21.79° around the [0001] axis [7]. Fig. 1(a) illustrates the constructed hcp–Y Σ7(0001) twist GB model, which consists of 12 atomic layers × 7 atoms per layer, totaling 84 Y atoms, with lattice parameters $a = b = 9.61$ Å, $c = 34.5$ Å, $α = β = 90°$, and $γ = 60°$. Due to the use of periodic boundary conditions, two equivalent grain boundaries are present in the model. The red dashed line I represents the GB interface between the upper and lower grains (denoted as the GB layer), while the green dashed line II corresponds to the interface between atomic layers closest to the boundary within the lower grain (denoted as the sub–GB layer). After full structural relaxation, the atomic configuration near the GB region (Layer I) exhibits significant displacement and distortion compared with the ideal bulk, whereas the structural distortion in the sub–GB region (Layer II) is relatively weak, though interlayer spacing changes are still observed. To systematically evaluate solute segregation behavior, two categories of segregation sites were investigated—14 substitutional and 11 interstitial sites. As shown in Fig. 1(a), seven inequivalent substitutional sites (labeled 1–7) are located within the GB and sub–GB layers. For the interstitial sites, illustrated in Fig. 1(b), the GB interstitial sites are located between layers within the GB region (Layer I), including octahedral sites (I) and tetrahedral sites (II), while the

sub–GB interstitial sites are situated in the sub−GB region (Layer II), comprising octahedral sites (III) and tetrahedral sites (IV). Fig. 1(c) presents the two free (0001) surfaces obtained by cleaving along the red dashed line I, with an approximately calculated surface energy of 0.99 J·m$^{-2}$, in excellent agreement with previously reported results [30,45]. These results confirm that the constructed GB model and computational parameters possess good accuracy and reliability, thereby providing a solid foundation for subsequent investigations of segregation and co–segregation behavior.

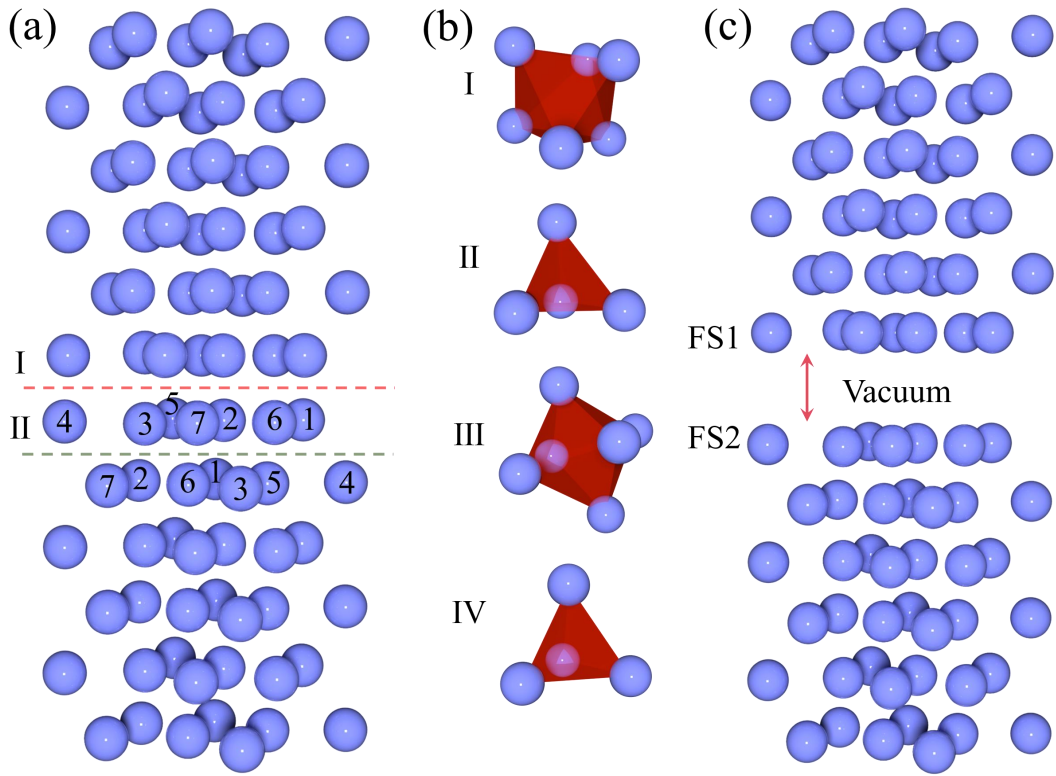

**Figure 1.** (a) The hcp–Y Σ7(0001) twist grain boundary (GB) model. The red dashed line I indicates the GB interface (GB plane) between the upper and lower grains, while the green dashed line II marks the interfacial layer between atomic planes closest to the boundary within the lower grain (sub–GB layer). Arabic numerals 1–7 denote the pre−segregation substitutional sites within one atomic layer.
(b) Configurations of interstitial sites in the GB region. I(Oct) and II(Ter) represent octahedral

and tetrahedral interstitial sites located between the GB planes (Layer I) shown in (a), respectively, while III(Oct) and IV(Ter) correspond to octahedral and tetrahedral interstitial sites in the sub–GB region (Layer II, green dashed line). "Oct" and "Ter" refer to octahedral and tetrahedral sites, respectively, giving a total of 11 interstitial sites.

(c) Two free (0001) surfaces formed by cleaving along the red line I.

## 3. Results and Discussion

### 3.1 Σ7(0001) Twist Grain Boundary vs. Ideal Bulk

As a common structural defect in materials, grain boundaries exhibit atomic arrangements that significantly deviate from those of the ideal bulk structure, often serving as preferential sites for crack initiation under external loading. The energy difference ($\Delta E_{Sep}$) between the two cleavage paths was calculated to evaluate the fracture behavior of the grain boundary. A negative $\Delta E_{Sep}$ indicates that the fracture along path I requires more external energy input than that along path II. As expressed in Eq. (1), $E_{GB}$ represents the total energy of the unfractured grain boundary structure, while $E_{FS1}^i$ and $E_{FS2}^i$ denote the total energies of the two free surfaces generated by cleavage along path $i$ ($i$ = I, II).

$$\Delta E_{Sep} = (E_{GB} - E_{FS1}^{I} - E_{FS2}^{I}) - (E_{GB} - E_{FS1}^{II} - E_{FS2}^{II}) \qquad (1)$$

The calculated $\Delta E_{Sep}$ value is 1.29 eV, indicating that fracture along path I requires less energy than that along path II. This result suggests that the grain boundary is more likely to fracture along path I, which may represent a potential crack initiation path for the Σ7(0001) boundary. Furthermore, to systematically compare the mechanical responses of the grain boundary and the bulk during tensile deformation, first-principles tensile simulations were performed to trace the energy evolution of the systems along the [0001] direction from the initial separation to complete fracture.

The first−principles tensile testing method has been widely applied to investigate fracture behaviors in grain boundaries, interfaces, and related systems. This approach involves gradually increasing the interlayer separation along a predefined fracture path, combined with either structural relaxation ("relaxed separation") or static total energy calculations ("rigid separation"), to obtain the energy evolution curve throughout the fracture process [46,47]. In this work, the rigid separation approach was adopted, with 16 different separation distances (0.25, 0.5, 0.75, 1.0, 1.5, 2.0, 2.5, 3.0, 3.5, 4.0, 4.5, 5.0, 5.5, 6.0, 7.0, and 8.0 Å) considered. The total energy of the system at each separation distance was calculated, and the separation energy was then obtained using Equation (2), where $E_x$ and $E_i$ are the total energies of the system at separation distance x and at equilibrium, respectively, and $S$ is the cross−sectional area in the xy plane. The resulting $E_{Sep}^x$ curve quantitatively characterizes the entire tensile process, providing a direct basis for comparing the cohesive strength between the grain boundary and the bulk [31,48].

$$E_{Sep}^x = (E_x - E_i)/S \qquad (2)$$

To compare the fracture response characteristics of the grain boundary and bulk structures during tensile deformation, different separation distances were introduced as pre−cracks along path I and path II of the Σ7(0001) twist grain boundary, as well as along the [0001] direction of hcp–Y bulk. The evolution of the separation energy with respect to the separation distance was then calculated under both relaxed and rigid separation conditions, as shown in Fig. 2.

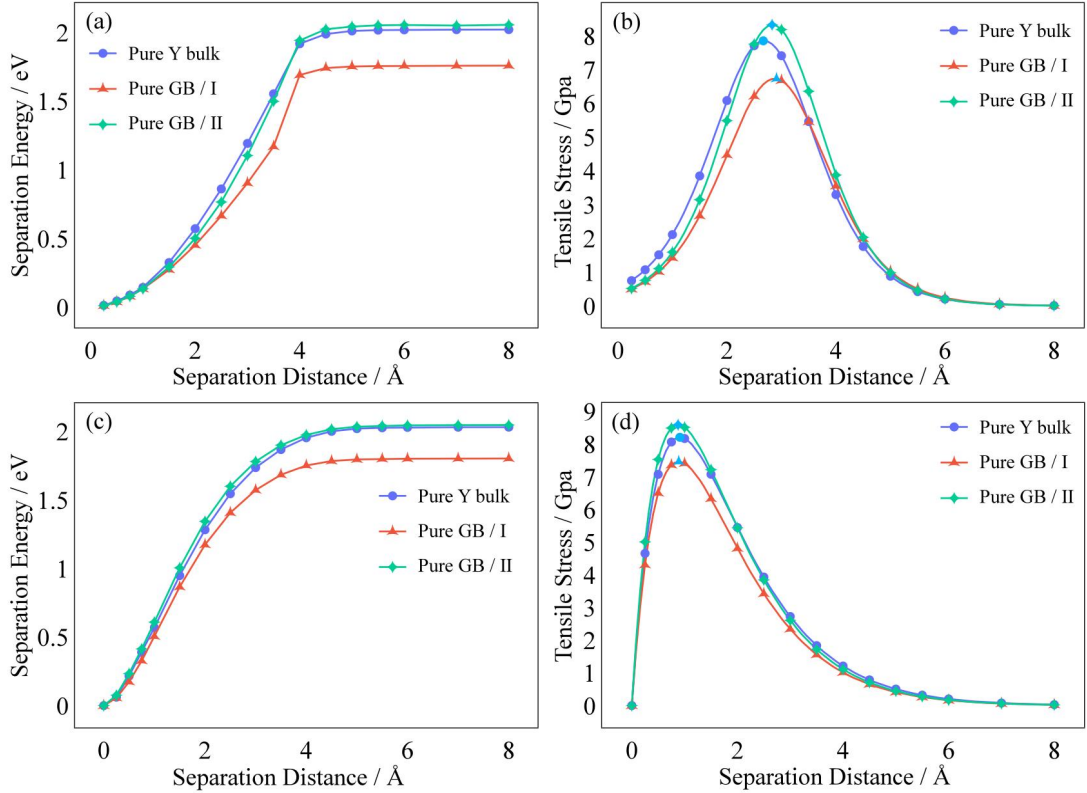

**Figure 2.** First−principles tensile simulation results for metallic Y bulk and the Σ7(0001) grain boundary under different fracture paths.
(a) and (b) Separation energies (a) and their derivatives (b) under relaxed separation conditions.
(c) and (d) Separation energies (c) and their derivatives (d) under rigid separation conditions.
Here, "Pure Y bulk" refers to the bulk Y results, while "Pure GB / I" and "Pure GB / II" correspond to the fracture modes of the grain boundary along path I and path II, respectively.

Fig. 2(a) and 2(c) present the separation energy curves under relaxed and rigid separation conditions, respectively, with the blue, red, and green curves representing the bulk, path I, and path II responses. As the separation distance $x$ increases, the system's separation energy gradually rises and stabilizes at approximately $x = 4$ Å, indicating complete fracture and the formation of two free surfaces. Throughout the tensile process, path I of the grain boundary consistently exhibits the lowest separation energy, whereas path II shows separation energies close to that of the bulk over a wide range of separation distances and remains higher than path I. This indicates that path II possesses greater fracture resistance, while path I represents the mechanically weakest region of the Σ7(0001) grain boundary. The energy required per unit area to fracture the bulk is slightly lower than that for path II but significantly higher than that for path I, suggesting that the twist grain boundary exhibits a certain degree of mechanical weakening relative to the ideal bulk, with the weakening primarily concentrated along path I.

It is noteworthy that the relationship between separation energy and separation distance exhibits a pronounced nonlinearity. The varying growth rates of the curve at different stages reflect the complex atomic rearrangements during the tensile process. The first derivative of the separation energy with respect to the separation distance can be regarded as the system's equivalent stress, representing the stress–strain response characteristics during tension, and can be used to further quantify the fracture initiation points and maximum load–bearing capacity of different structures.

Under relaxed separation conditions, the separation energy–distance curves were numerically fitted, and their first derivatives were calculated to obtain the corresponding tensile stress–separation distance relationships, as shown in Fig 2(b). For rigid separation calculations, the statically converged separation energy–distance curves were fitted using the Rose equation, as expressed in Equation (3) [49,50]. In this equation, $\lambda$ represents the characteristic separation distance, and $E_{frac}$, defined by Equation (4), denotes the separation energy as $x \to \infty$. By differentiating the Rose equation, $f'(x)$ can be obtained, as shown in Equation (5), which describes the stress–strain relationship under rigid separation conditions. The corresponding results are plotted in Fig 2(d), where the stress peak at $x = \lambda$ indicates the fracture critical point corresponding to the maximum tensile capacity of the system.

$$f(x) = E_{frac} - E_{frac}(1 + x/\lambda) e^{(-x/\lambda)} \qquad (3)$$

$$E_{frac} = \lim_{x \to \infty} E_{Sep}^{x} = (E_{\infty} - E_{GB})/S \qquad (4)$$

$$f'(x) = e^{(-x/\lambda)} E_{frac} x / \lambda^2 \qquad (5)$$

As shown in Fig. 2(b) and 2(d), the tensile stress gradually increases with increasing separation distance, reaches a peak at a specific separation, and then rapidly decreases. In the early stage of tension, strong interactions exist between atoms on either side of the fracture path, so the system must overcome significant chemical bond forces, resulting in a continuous rise in stress. As the separation distance further increases, bond breaking leads to a significant reduction in interatomic forces, and the resistance the system must overcome sharply decreases, causing stress to rapidly decay to zero, corresponding to complete fracture. It is noteworthy that the peak stress positions under relaxed and rigid separation do not completely coincide. In relaxed separation, atoms are allowed to fully relax at each step, and some atoms can locally rearrange to release internal stress, thereby delaying fracture and shifting the stress peak to a larger separation distance.

In first–principles tensile simulations, the maximum tensile stress that the system can sustain is defined as the ideal tensile strength, representing the intrinsic strength limit of the material. As shown in Fig. 2, under both relaxed and rigid separation, the strength order of the three systems remains consistent: grain boundary path II is slightly higher than bulk Y, and both are significantly higher than grain boundary path I. This trend agrees with the trend observed from the fracture energy difference analysis of the two cleavage paths [31,47,51,52], indicating that fracture along path I not only requires the least external energy but also exhibits the lowest ideal tensile strength, confirming it as the preferred fracture path of the Σ7(0001) twist grain boundary. These results suggest that the Σ7(0001) twist grain boundary exhibits a certain degree of mechanical weakening compared with the ideal bulk, with the weakening primarily concentrated along path I.

**3.2 Effects of Single−Element Solute Segregation Near the Grain Boundary**

**3.2.1 Segregation Tendencies of Single−Element Solutes**

"Grain boundary segregation engineering" is an effective strategy to optimize material properties by controlling the enrichment behavior of solute elements at grain boundaries. It has been widely applied to enhance key alloy characteristics such as mechanical strength and thermal stability [11,53]. To determine whether a solute element has a thermodynamic driving force to segregate to the grain boundary region, the segregation energy is commonly used as a criterion. The segregation energy is defined as the difference between the dissolution energy of the solute near the grain boundary and that in the ideal bulk, as expressed in Equation (6). When $E_{Seg}^{GB/x} < 0$, the migration of solute atoms from the bulk to the grain boundary lowers the total energy of the system, indicating a thermodynamically spontaneous process. Conversely, if $E_{Seg}^{GB/x} > 0$, the solute does not tend to segregate to the grain boundary. To calculate the dissolution energies of solutes in different structural environments, two doping mechanisms are considered: substitutional doping and interstitial doping. Their respective dissolution energies are given by Equations (7) and (8).

$$E_{Seg}^{GB/x} = E_{disolve}^{GB/x} - E_{disolve}^{Bulk/x} \quad (6)$$

$$E_{disolve}^{GB/x} = E_{GB/x} - E_{GB} - E_x \quad (7)$$

$$E_{disolve}^{GB/x} = E_{GB/x} - (E_{GB} - E_Y) - E_x \quad (8)$$

In the equations, $E_{disolve}^{GB/x}$ and $E_{disolve}^{Bulk/x}$ represent the dissolution energies of the solute element at the grain boundary and in the bulk, respectively. $E_{GB/x}$ and $E_{GB}$ denote the total energies of the grain boundary containing the solute and the pure grain boundary, while $E_Y$ and $E_x$ correspond to the average atomic energies of metallic Y and the solute element in their respective bulk phases. All energy values are obtained from fully relaxed structures. For brevity, the detailed dissolution energies of all elements in the grain boundary and bulk are provided in Table S1 of the Supplementary Material.

As described in Section 2, due to the periodicity of the computational model, the grains above and below the grain boundary are structurally equivalent. Therefore, only the lower grain is considered as the representative region for analyzing solute segregation behavior [47]. The atomic layers indicated by the red and green dashed lines in Fig. 1 are defined as Layer I and Layer II, serving as potential substitutional doping layers for the solute elements. Each layer contains seven possible substitutional sites, totaling fourteen candidate segregation positions.

In previous studies on symmetrical twin boundaries, the grain boundary structure exhibits mirror symmetry, and the atomic environments within the same layer are usually equivalent. Hence, only a representative site needs to be selected to describe segregation trends [31]. However, in the Σ7(0001) twist grain boundary considered in this study, the upper and lower grains are rotated by a certain angle around the [0001] axis, resulting in a loss of mirror symmetry at the grain boundary, and atoms within the same layer are no longer equivalent. To comprehensively evaluate the variation of segregation energies at different positions, all possible substitutional sites in Layer I and Layer II are systematically examined and numbered as 1/I–7/I and 1/II–7/II, respectively, enabling a comparative analysis of the energetic stability and segregation tendency at each site in subsequent discussions.

Fig. 3 presents the segregation energy distributions of eleven solute elements (Al, Cr, Cu, Fe, Mg, Mo, Nb, Ti, Zn, Zr, and Si) at the different substitutional sites (1/I–7/I and 1/II–7/II) of the Σ7(0001) twist grain boundary (numerical values are provided in Tables S2 and S3 of the Supplementary Material). It is evident that the segregation behavior of different solute elements at the grain boundary exhibits significant variation. Cu, Fe, Zn, and Si show relatively smooth changes in segregation energy across the sites in Layer I, displaying similar trends, whereas other elements (e.g., Al, Cr, Mg, Mo, Nb, Ti, Zr) exhibit more pronounced differences in segregation energy among different sites, indicating a stronger influence of the local structural environment on their segregation tendencies. Overall, elements with negative segregation energies possess a thermodynamic driving force to enrich at the grain boundary. In the systems studied, most solute elements, except Fe, exhibit a certain degree of segregation tendency. Specifically, nine elements (Al, Cr, Mg, Mo, Nb, Ti, Zn, Zr, and Si) show significant negative segregation energies at substitutional sites within the grain boundary, while Cu shows segregation only at sites in Layer I. In contrast, Fe exhibits positive segregation energies at all sites, indicating a lack of thermodynamic driving force for segregation at the grain boundary. Among all substitutional doping sites, the lowest segregation energies for each element occur at the following sites: Al at 1/I, Cr at 3/I, Cu at 7/I, Fe at 7/I, Mg at 1/I, Mo at 3/I, Nb at 3/I, Ti at 3/I, Zn at 3/I, Zr at 6/II, and Si at 2/I. The corresponding energies are −0.260 eV, −0.384 eV, −0.260 eV, 0.645 eV, −0.271 eV, −0.299 eV, −0.419 eV, −0.259 eV, −0.458 eV, −0.232 eV, and −0.531 eV, respectively.

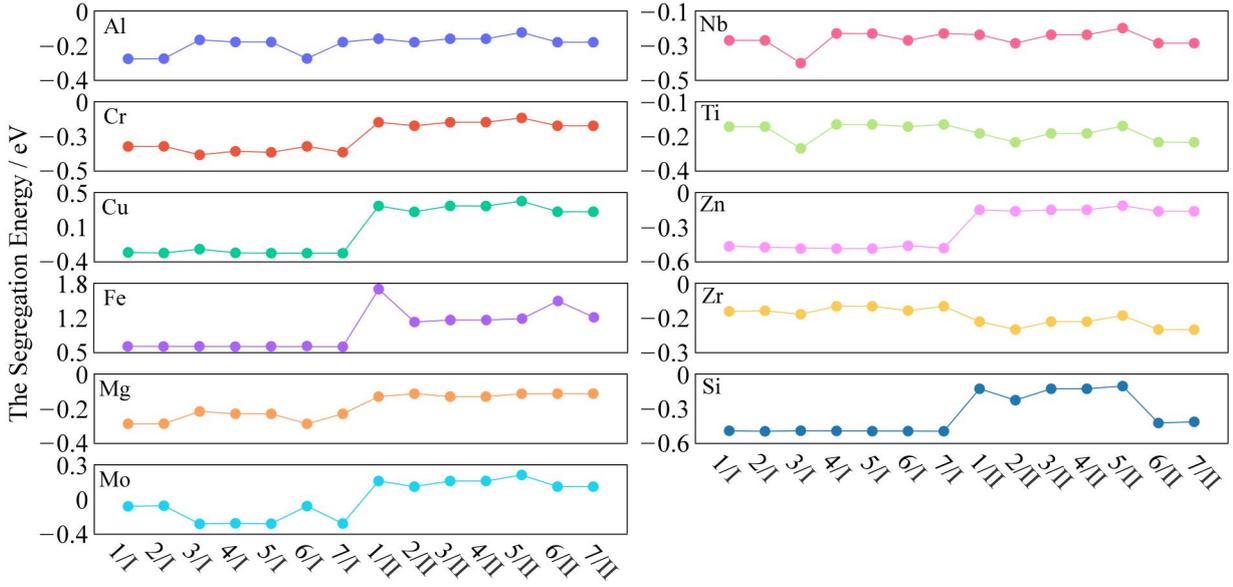

**Figure 3.** Segregation energy distributions of eleven low−neutron−absorption metallic elements at substitutional sites in Layer I and Layer II near the Σ7(0001) twist grain boundary. The notation *n*/I and *n*/II denotes the *n*–th (1–7) potential segregation site in Layer I and Layer II, respectively.

In previous studies on symmetric grain boundaries [29,54], due to the relatively small differences in atomic radii among metallic elements, research typically focused only on substitutional segregation of metal solutes at the grain boundary, while interstitial sites were primarily considered as segregation sites for nonmetallic elements. However, the Σ7(0001) grain boundary region in hcp–Y contains relatively large tetrahedral and octahedral interstitial sites, which can provide stable embedding spaces for specific elements. Therefore, in this study, we further included the octahedral and tetrahedral interstitial sites near the Y Σ7(0001) twist grain boundary (schematically shown as sites I, II, III, and IV in Fig. 1(b)) in the segregation analysis. This allows for a systematic evaluation of both substitutional and interstitial segregation trends of solute atoms in the grain boundary region, thereby providing a more comprehensive understanding of their segregation characteristics.

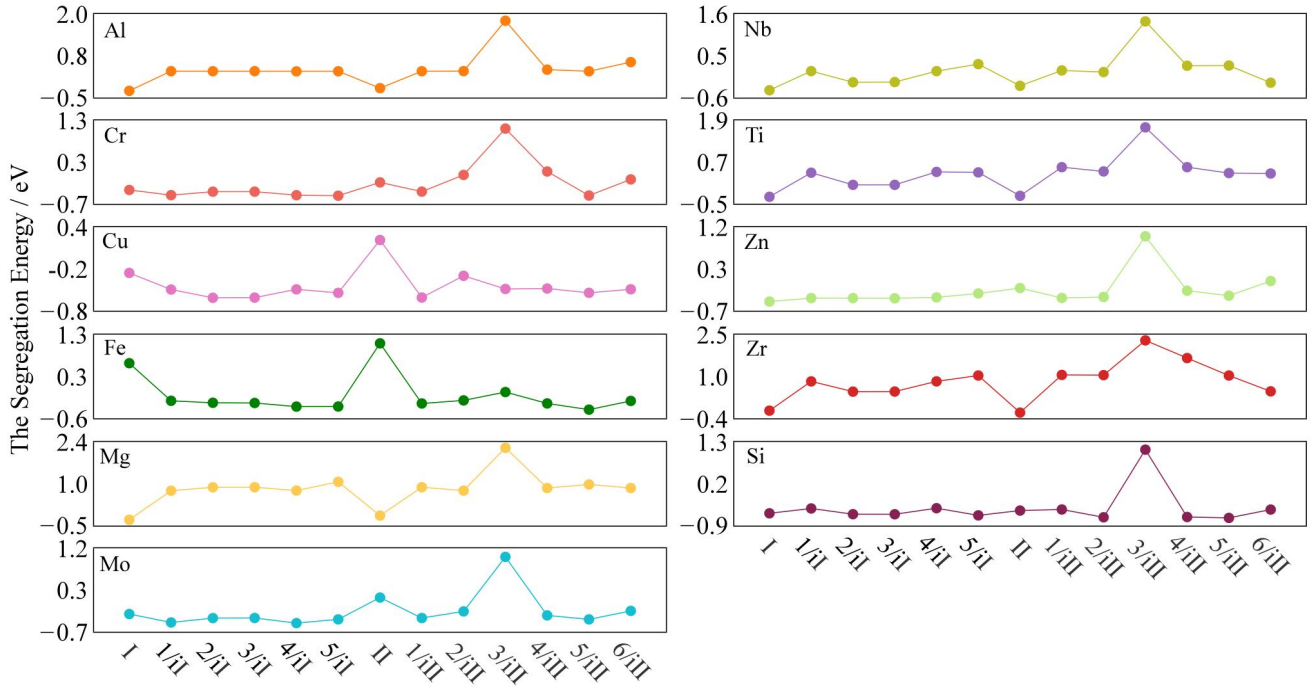

**Figure 4.** Segregation energies of solute elements at preselected sites near the grain boundary. The x−axis labels I and II indicate the optimal substitutional segregation sites in Layers I and II, respectively. Labels n/iI and n/iII denote the nth (1–6) preselected interstitial segregation site in Layers I and II, where *i* indicates an interstitial position.

Fig. 4 presents the lowest segregation energy sites for each solute element in Layers I and II near the Σ7(0001) twist grain boundary, as well as their segregation trends at various interstitial sites (detailed segregation energy values are provided in Table S4 of the Supplementary Material). The results reveal significant differences in the segregation behavior of different elements between interstitial and substitutional sites, highlighting clear element−dependent thermodynamic characteristics. Fe exhibits the most distinctive behavior: its segregation energies at all substitutional sites are positive, while at interstitial sites, negative segregation energies are observed, with the lowest energy at the 5/iI site. This indicates that Fe segregation is primarily driven by the grain boundary interstitials. In contrast, Al, Mg, Ti, and Zr show positive segregation energies at interstitial sites, suggesting that these metal solutes predominantly occupy substitutional positions at the grain boundary, with minimal interstitial segregation tendency. Nb exhibits a slightly negative segregation energy at the III site (Octahedral interstice, Oct−i), while all other interstitial sites are positive, indicating a limited segregation drive only in the local octahedral environment, with substitutional segregation remaining the dominant mode. Cr, Cu, Mo, Zn, and Si show pronounced segregation tendencies at interstitial sites. Specifically, the minimum segregation energies of Cr, Cu, Mo, and Zn at interstitial sites are lower than at substitutional sites in Layers I and II, indicating higher stability within the grain boundary interstitial structures. Notably, Si exhibits the lowest interstitial segregation energy (−0.618 eV), demonstrating a strong adsorption capability at the grain boundary. This behavior is closely related to its small atomic radius and covalent bonding ability, allowing it to stably occupy tetrahedral interstitial sites in the Y grain boundary. A comprehensive comparison of optimal segregation sites for all solute elements indicates that Al, Mg, Ti, Nb, and Zr preferentially occupy substitutional sites replacing Y atoms, whereas Cr, Cu, Mo, Zn, Fe, and Si are more inclined toward interstitial sites at the grain boundary. The specific optimal segregation sites are as follows: Al (1/I), Cr (5/iI, octahedral interstice), Cu (2/iI, octahedral interstice), Fe (5/iII, tetrahedral interstice), Mg (1/I), Mo (4/iII, tetrahedral interstice), Nb (3/I), Ti (3/I), Zn (3/I), Zr (6/II), and Si (5/iII, tetrahedral interstice). Ranking the segregation driving forces from strongest to weakest based on segregation energy: Si (−0.650 eV) > Cu (−0.643 eV) > Cr (−0.524 eV) > Mo (−0.510 eV) > Zn (−0.458 eV) > Fe (−0.424 eV) > Nb (−0.419 eV) > Mg (−0.271 eV) > Al (−0.260 eV) > Ti (−0.259 eV) > Zr (−0.232 eV).

**3.2.2 Effect of Elemental Segregation on Grain Boundary Stability**

Grain boundaries are ubiquitous defects in materials, significantly influencing their mechanical properties, thermal stability, and diffusion behavior. The grain boundary energy ($\gamma$) is a key parameter characterizing the thermodynamic stability of a grain boundary, reflecting the energetic instability of the boundary structure relative to the ideal bulk. A lower grain boundary energy corresponds to a more stable boundary. The grain boundary energy can be defined as the energy required to form a grain boundary from an ideal bulk structure, normalized by the unit area (J·m$^{-2}$). The grain boundary energy of a pure boundary is calculated as follows [29–31]:

$$\gamma_{GB/x} = ( E_{GB} - N_{GB} \times E_Y ) / ( 2S ) \qquad (9)$$

For a grain boundary doped with solute elements, the formula is given as:

$$\gamma_{GB/x} = ( E_{GB}^x - N_{GB}^x \times E_Y - E_x ) / S - \gamma_{GB} \qquad (10)$$

The symbols in the equation are defined as follows: $\gamma_{GB}$ represents the grain boundary energy of the pristine boundary; $N_{GB}$ and $N_{GB}^x$ denote the number of Y atoms in the clean and solute−doped grain boundary models, respectively, and $S$ is the area of the $xy$ plane of the grain boundary model. To investigate the effect of solute segregation on grain boundary stability, this study calculated the grain boundary energy for each element based on its optimized segregation site. The stabilization effect of solute segregation was evaluated by comparing the energy difference between the doped and pristine grain boundaries, defined as $\Delta\gamma = \gamma_{GB/x} - \gamma_{GB/x}$. A negative $\Delta\gamma$ indicates that solute segregation lowers the grain boundary energy and enhances its thermodynamic stability, whereas a positive $\Delta\gamma$ suggests that doping reduces the stability of the grain boundary.

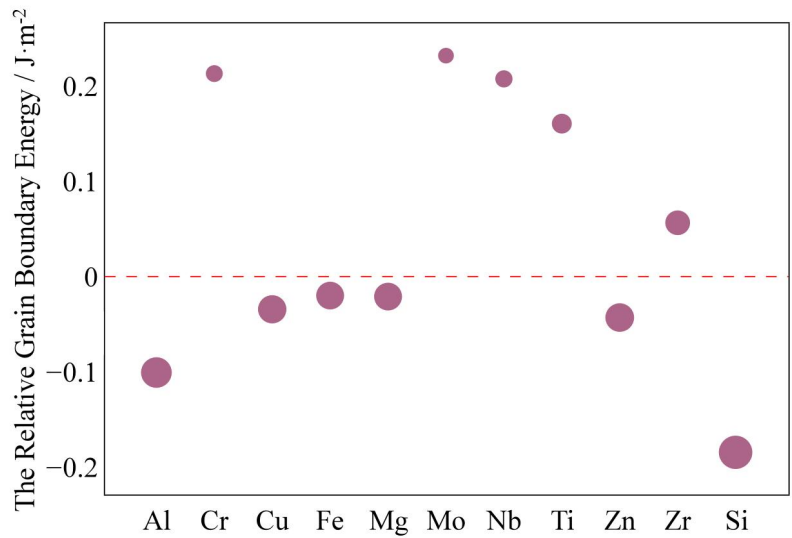

**Figure 5**. Grain boundary energies of various elements at their most favorable segregation sites.

Fig. 5 shows the relative changes in grain boundary energy ($\Delta\gamma$) for each solute element at their optimal segregation sites, with the red dashed line representing $\Delta\gamma = 0$. The size of each data point is proportional to the magnitude of the grain boundary energy reduction, meaning that larger points indicate stronger stabilization effects due to solute segregation. The results indicate that certain elements weaken grain boundary stability: Cr, Mo, Nb, Ti, and Zr all increase the grain boundary energy ($\Delta\gamma > 0$), suggesting a destabilizing effect. Among these, Zr has the weakest impact, whereas Mo causes the most significant increase, indicating a strong destabilization tendency. Conversely, solute elements that enhance grain boundary stability—Al, Cu, Fe, Mg, Zn, and Si—reduce the grain boundary energy ($\Delta\gamma < 0$), indicating a stabilizing effect. Among them, Fe and Mg provide relatively weak stabilization, while Si exhibits the most pronounced reduction in grain boundary energy, suggesting the strongest grain boundary strengthening tendency. In terms of magnitude of stabilization, the order is: Si > Al > Zn > Cu > Mg > Fe, with Δγ values of −0.184, −0.101, −0.043, −0.034, −0.021, and −0.020 J·m$^{-2}$, respectively. For elements that destabilize the grain boundary, the order is: Zr < Ti < Nb < Cr < Mo, with Δγ values of 0.056, 0.160, 0.207, 0.213, and 0.232 J·m$^{-2}$, respectively.

### 3.2.3 Segregation−Induced Grain Boundary Strengthening and Embrittlement

As discussed in Section 3.1, the theoretical tensile strength of the Σ7(0001) twist grain boundary is significantly lower than that of the ideal Y bulk, with path I identified as the preferential fracture path of the boundary. Therefore, subsequent analyses of the mechanical influence of solute segregation on the grain boundary are all carried out along path I.

To quantitatively describe the strengthening or embrittlement tendency induced by solute segregation, the grain boundary strengthening energy ($E_{Strength}^{GB/x}$) based on the Rice–Wang model is employed [23,29–31,55]. This parameter is defined as the difference between the fracture energy of the solute–segregated boundary and that of the clean boundary. A negative strengthening energy indicates that more energy is required to fracture the segregated boundary compared to the pristine one, implying that solute segregation enhances grain boundary strength. Conversely, a positive value signifies a weakening (embrittlement) effect. The strengthening energy is calculated using Eq. (11).

$$E_{Strength}^{GB/x} = ( E_{GB/x} - E_{GB} ) - ( E_{FS/x} - E_{FS} ) \qquad (11)$$

Fig. 6 presents the strengthening energies of various solute elements at their most favorable segregation sites, where the size of each circle represents the degree of strengthening (a more negative strengthening energy indicates a stronger strengthening effect). The calculated results reveal that Al, Mg, Zn, and Zr exhibit positive strengthening energies, indicating that their segregation weakens the interfacial bonding strength and thus induces embrittlement. Among them, Mg shows the highest strengthening energy (0.181 eV), implying the most pronounced weakening effect on the grain boundary strength. The embrittlement tendency follows the order: Mg > Zn > Zr > Al, with corresponding strengthening energies of 0.181 eV, 0.172 eV, 0.097 eV, and 0.078 eV, respectively. In contrast, Mo, Fe, Si, Cr, Cu, Nb, and Ti exhibit negative strengthening energies, indicating that their segregation increases the energy required for interfacial fracture, thereby enhancing the grain boundary strength. Among these, Mo exhibits the strongest strengthening effect, reducing the grain boundary strengthening energy by 1.603 eV compared with the pristine interface, followed by Fe (−1.421 eV). The overall strengthening trend can be ranked as: Mo > Fe > Si > Cr > Cu > Nb > Ti, with corresponding strengthening

energies of −1.603 eV, −1.421 eV, −0.706 eV, −0.374 eV, −0.354 eV, −0.293 eV, and −0.210 eV, respectively.

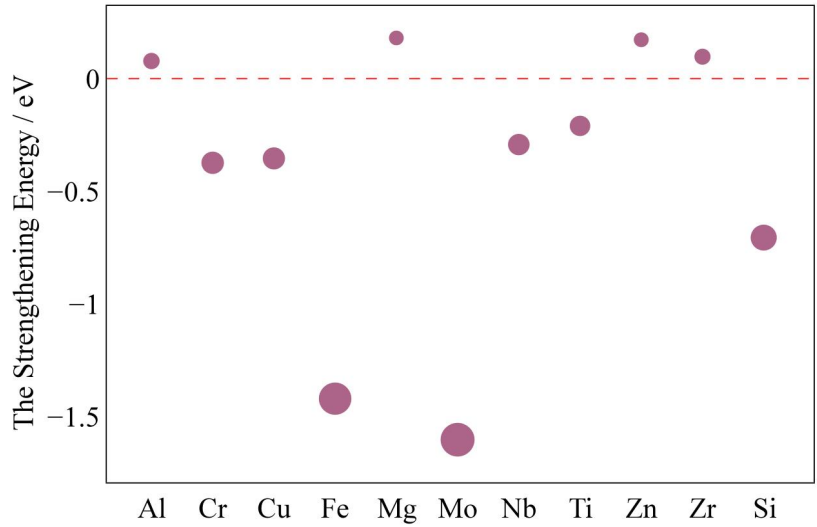

**Figure 6**. Strengthening energies of various solute elements at their most favorable segregation sites.

To further elucidate the synergistic influence of solute segregation on grain boundary stability and strength, Fig. 7 presents the relationship between the relative grain boundary energy ($\Delta\gamma$) and the strengthening energy. In the figure, the x−axis represents the strengthening energy, while the y−axis represents the relative grain boundary energy. A leftward red arrow indicates grain boundary strengthening ($E_{Strength}^{GB/x} < 0$), whereas a downward arrow denotes enhanced thermodynamic stability ($\Delta\gamma < 0$). The shaded region corresponds to solute elements that simultaneously strengthen and stabilize the grain boundary. As shown in Fig. 7, the eleven solute elements exhibit four distinct regulatory behaviors:

Strengthening but destabilizing (upper left quadrant): Mo, Cr, Nb, and Ti can significantly enhance the grain boundary strength but at the cost of increased boundary energy. This implies that, while their segregation benefits mechanical performance, it thermodynamically destabilizes the boundary ($\Delta\gamma > 0$).

Embritting and destabilizing (upper right quadrant): Zr exhibits adverse effects in both aspects, weakening the grain boundary strength and decreasing its thermal stability ($E_{Strength}^{GB/x} > 0$ and $\Delta\gamma > 0$).

Embritting but stabilizing (lower right quadrant): Mg, Zn, and Al reduce the grain boundary energy ($\Delta\gamma < 0$), thereby improving thermodynamic stability, but concurrently cause embrittlement ($E_{Strength}^{GB/x} > 0$).

Strengthening and stabilizing (lower left shaded quadrant): Fe, Cu, and Si fall within this ideal regime, showing both strengthening and stabilizing effects. Among them, Si stands out with the most remarkable overall performance—its segregation yields the greatest reduction in grain boundary energy while providing a pronounced strengthening effect.

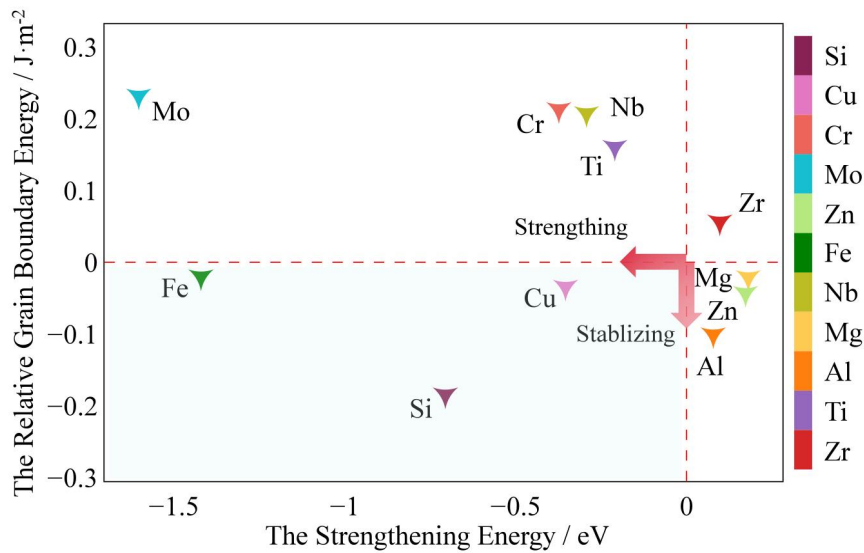

**Figure 7**. Scatter plot of the strengthening energy ($x$−axis) versus the relative grain boundary (GB) energy ($y$−axis) for various elements after segregation.

### 3.3 Co−segregation Effects of Si+$x$ Solute Elements near the Grain Boundary

As shown in Fig. 8(a), when a Si atom is initially placed at the pre−segregation site 5/iII (a tetrahedral interstitial site), it migrates across the Y atomic layer and finally stabilizes at an interstitial position near the grain boundary interface (Layer I) after structural relaxation, resulting in a new energetically favorable configuration. This configuration exhibits both enhanced grain boundary stability and a pronounced strengthening effect. Therefore, based on this stable Si pre−segregated grain boundary, the present study further investigates the co−segregation behavior of other low thermal neutron absorption cross−section elements (Al, Cr, Cu, Fe, Mg, Mo, Nb, Ti, Zn, and Zr) within the same system. To this end, six Y atomic sites adjacent to the segregated Si atom were selected as potential co−segregation positions, as illustrated in Fig. 8(a). Each solute element was individually placed at these positions to calculate its corresponding segregation energy. The segregation energy in the Si−presegregated grain boundary system is defined as follows in Eq. (12), where $E_{disolve}^{GB/Si+x}$ represents the dissolution energy of solute element $x$ in the Si−presegregated grain boundary configuration.

$$E_{Seg}^{Si+x} = E_{disolve}^{GB/Si+x} - E_{disolve}^{Bulk/x} \qquad (12)$$

Fig. 8(b) illustrates the variation trends of segregation energies for different solute elements in the Si pre−segregated grain boundary. All segregation energies are found to be negative, indicating that these elements are thermodynamically inclined to enrich in the vicinity of the Si atom at the grain boundary, demonstrating a pronounced co−segregation driving force. The small differences in segregation energies among various sites suggest that the attractive effect of Si on other solutes is relatively uniform within the local region. Among these, site 2 exhibits the lowest segregation energy, representing the most favorable co−segregation position for all elements in this system. As shown in Table 1, compared with the clean grain boundary, the pre−segregation of Si significantly enhances the segregation tendency of other solutes, as evidenced by the further reduction of their optimal segregation energies. This indicates that the presence of Si facilitates the migration of solute atoms toward the grain boundary. Among all elements, Mo exhibits the strongest segregation tendency, with a segregation energy of −1.241 eV. Ti shows the most significant enhancement in segregation tendency, with its segregation energy decreasing from −0.259 eV to −0.841 eV, corresponding to a 224.7% increase, revealing a strong co−segregation driving effect. In contrast, Fe shows the smallest enhancement, with its segregation energy reduced by only −0.134 eV, corresponding to a 31.6% increase in segregation tendency. The order of enhancement in segregation tendency induced by Si pre−segregation is as follows: Ti (224.7%) > Al (188.1%) > Nb (158.5%) > Mo (150.7%) > Zr (122%) > Zn (110%) > Cr (104.4%) > Cu (67.9%) > Mg (60.5%) > Fe (31.6%).

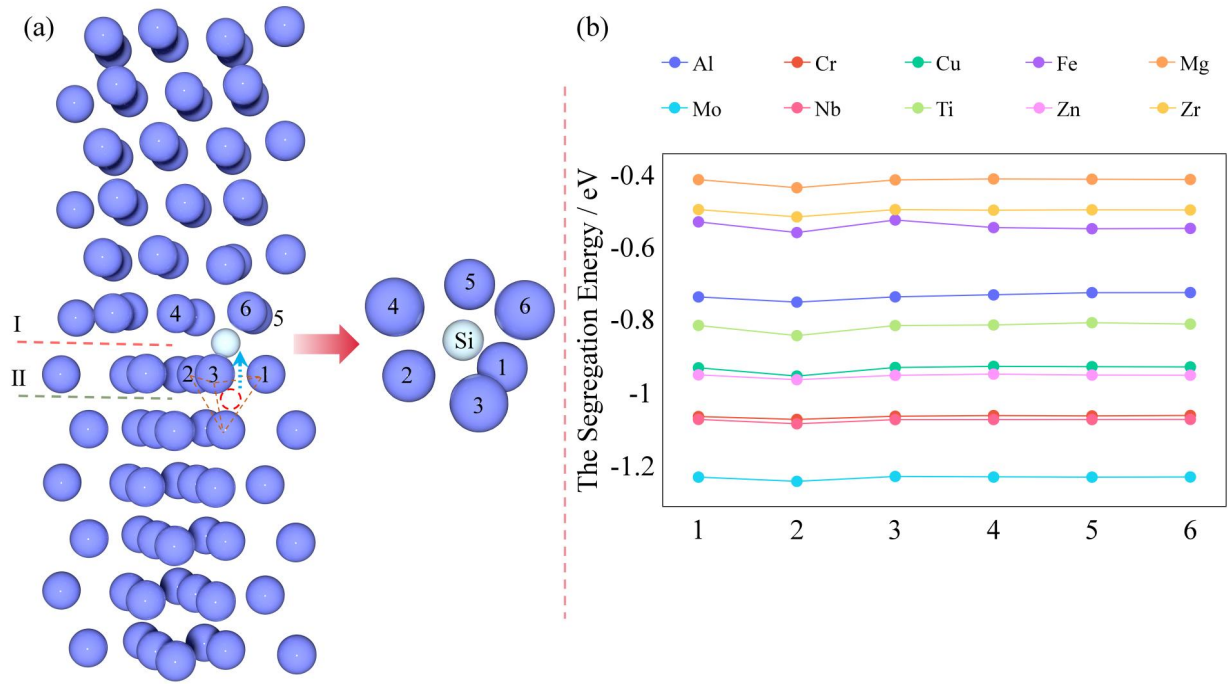

**Figure 8**. Si−induced co−segregation models of other solute elements at the grain boundary and the corresponding segregation energies of each element at the pre−segregated sites.

**Table 1**. Segregation energies (in eV) of various solute elements at their most favorable segregation sites in the Si pre−segregated grain boundary system and the clean grain boundary.

| Element | Co−Seg | Seg | Element | Co−Seg | Seg |
|---|---|---|---|---|---|
| Al | −0.749 | −0.260 | Mo | −1.241 | −0.495 |
| Cr | −1.071 | −0.524 | Nb | −1.083 | −0.419 |
| Cu | −0.952 | −0.567 | Ti | −0.841 | −0.259 |
| Fe | −0.558 | −0.424 | Zn | −0.962 | −0.458 |
| Mg | −0.435 | −0.271 | Zr | −0.515 | −0.232 |

* The "Co–Seg" column lists the segregation energies of each element in the Si pre−segregated grain boundary system.
* The "Seg" column lists the segregation energies of each element at the optimal segregation site in the pure grain boundary system.

Fig. 9 shows the relative grain boundary energy (a) and strengthening energy (b) of various solute elements at their most favorable co−segregation sites (see the Supplementary Material, Table S5 for numerical details). The purple dots correspond to systems with individual solute segregation ($x$), while the orange dots represent Si+$x$

co−segregation systems. As shown in Fig. 9(a), all Si+$x$ co−segregation systems exhibit lower relative grain boundary energies compared with their corresponding single−solute segregation systems, indicating a significant synergistic effect between Si and other solutes that further reduces grain boundary energy and enhances thermodynamic stability. Notably, for elements such as Cr, Mo, Nb, Ti, and Zr—which originally destabilize the grain boundary—their co−segregation with Si effectively offsets these adverse effects, leading to markedly improved stability. This suggests that Si acts as a "regulator," mitigating the instability induced by certain solutes and promoting overall boundary stabilization. As shown in Fig. 9(b), the co−segregation of Si+$x$ also plays an important role in determining grain boundary strength. For solutes that originally act as embrittlers, such as Al, Mg, Zn, and Zr, the pre−segregation of Si significantly alters their segregation−induced mechanical effects, transforming them into effective grain boundary strengtheners. Moreover, the strengthening effects of Nb and Ti are further enhanced by 106.5% and 212.4%, respectively. Conversely, for Cr, Cu, Fe, and Mo, co−segregation with Si weakens their strengthening effects, though this reduction is accompanied by notable improvements in boundary stability. In particular, the Si+Fe co−segregation system shows a substantial decrease in grain boundary strength, approaching that of the clean boundary, while the Si+Cr, Si+Cu, and Si+Mo systems exhibit strength reductions of 10.4%, 31.6%, and 53.1%, respectively.

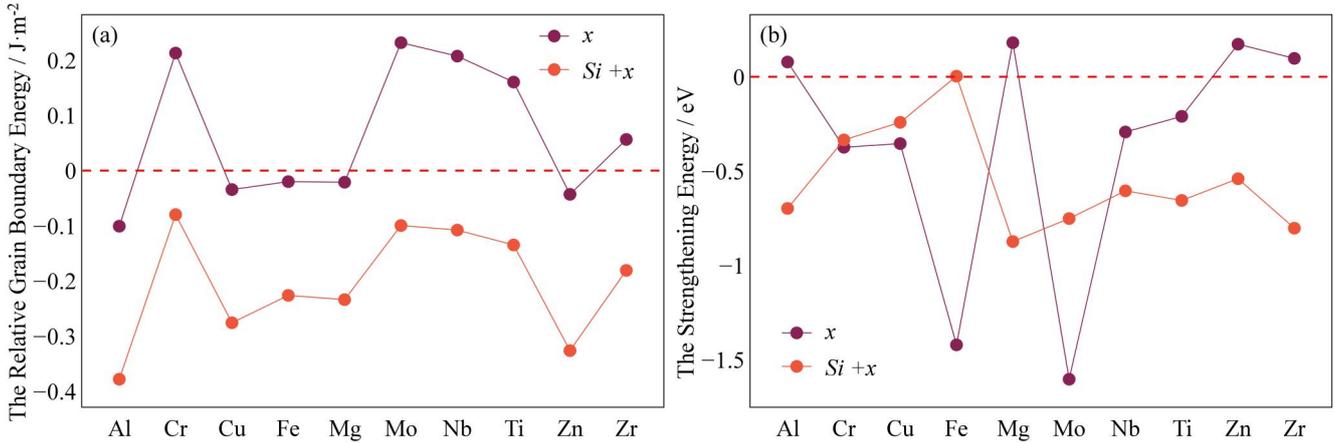

**Figure 9**. Relative grain boundary energy (a) and strengthening energy (b) of various solute elements at their most favorable co−segregation sites.

Fig. 10 presents the relationship between strengthening energy and relative grain boundary (GB) energy in the Si+$x$ co−segregation systems, where the intersection of the blue dashed lines (marked by brown upward triangles) denotes the reference value of the single−Si segregated GB. It can be seen that, except for the Si+Fe system, all Si+$x$ co−segregation systems fall into the lower−left quadrant, indicating a significant synergistic interaction between Si and these solute elements. Such synergy simultaneously reduces both GB energy and strengthening energy, thereby promoting the stabilization and strengthening of the GB. In the Si+Fe system, the strengthening energy is positive and the GB energy is slightly higher, suggesting that its mechanical performance is comparable to or slightly weaker than that of the pure GB, while maintaining relatively good thermodynamic stability. In contrast, the Si+Mg system exhibits the most pronounced effect, with both strengthening energy and GB energy substantially lower than those of the single−Si segregated system, indicating that the hybrid interaction between Mg and Si effectively enhances GB strength and structural stability. The Si+Al system shows the lowest GB energy, revealing the strongest stabilization tendency. Although its strengthening energy is slightly higher than that of the single−Si case, the overall bonding strength remains strong. For the Si+Zr system, the GB energy is nearly identical to that of the single−Si system, but the strengthening energy is considerably more negative, implying that Zr enhances GB strength without compromising interfacial thermal stability. This suggests that the metallic nature of Zr contributes to maintaining structural integrity while forming a complementary bonding interaction with Si. In the Si+Mo system, the strengthening energy decreases relative to the single−Si system, indicating an improvement in GB strength, whereas the slight increase in GB energy implies a modest destabilization accompanying the enhanced strength. For the Si+Ti, Si+Nb, and Si+Cr systems, both strengthening and GB energies exhibit partial increases, implying that these transition metals may engage in competitive interactions with Si. Such interactions could induce local charge redistribution or stress concentration, thereby weakening GB strength. Conversely, the Si+Cu and Si+Zn systems show higher strengthening energy but lower GB energy compared with the single−Si case, indicating that these solutes preferentially promote thermodynamic stabilization rather than mechanical strengthening.

Overall, the Si co−segregation systems display distinct synergistic behaviors with different solute elements. These findings demonstrate that by tailoring the co−segregation partners of Si, it is possible to achieve precise control over GB stability and strength at the atomic scale, providing a theoretical basis for grain boundary engineering design.

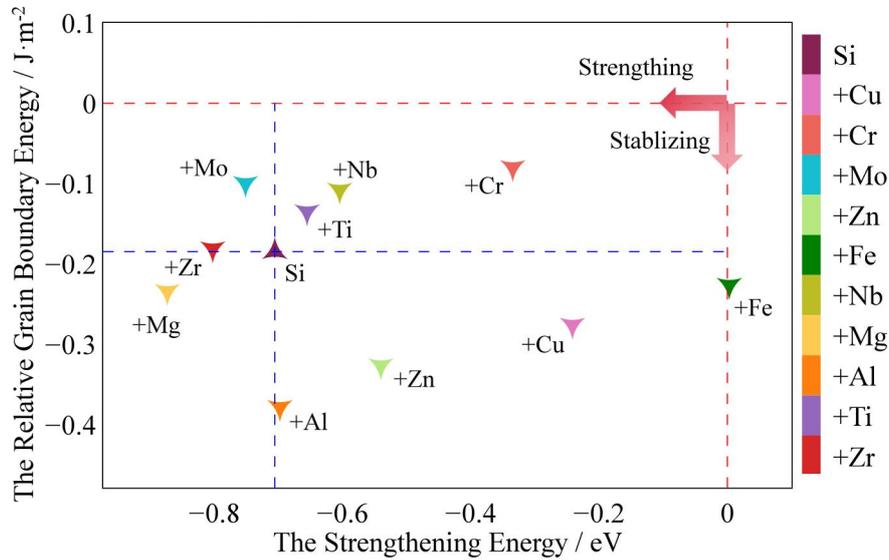

**Figure 10**. Scatter plot of strengthening energy (*x*−axis) and relative grain boundary energy (*y*−axis) for various Si+*x* co−segregation systems.

**3.4 Electron Localization Function and Density of States Analysis**

To further clarify the microscopic mechanisms responsible for the variation in grain boundary (GB) strength induced by elemental segregation, the electron localization function (ELF) of the Σ7(0001) twist GB in Y metal was analyzed under different segregation conditions, including the pure GB, single−Si segregation, and Si+Mg co−segregation systems, as illustrated in Fig. 11. The ELF isosurfaces were visualized using the VESTA software with an isosurface value of 0.64, providing an intuitive representation of the electron localization distribution in the GB region. This analysis offers valuable insight into the synergistic interactions between segregated elements and reveals how these interactions modulate the GB bonding characteristics and local electronic structures [56,57].

In the pure Y GB system ([Fig. 11(a)](#)), the interlayer electron distribution within the grains is relatively uniform, whereas the electron localization in the GB region (Layer I) is significantly weakened. This region is dominated by typical metallic bonding, where electrons are delocalized and exhibit weak bonding directionality, resulting in relatively low cohesive strength between GB atoms. Such weak electron localization indicates that the interatomic interactions mainly rely on metallic bonding mediated by free electrons, which are insufficient to resist bond rupture or crack propagation under external loading. This observation is consistent with the analysis of the preferential crack propagation path (Path I) discussed in Section 3.1. When a Si atom segregates into the GB ([Fig. 11(b)](#)), a remarkable enhancement in electron localization occurs within the GB region. Due to the higher electronegativity and smaller atomic radius of Si, it can occupy the interstitial sites between Y atoms and induce a redistribution of electron density along the Y–Si bonding direction. The apparent accumulation of electrons around Si indicates the formation of polar covalent bonds between Si and adjacent Y atoms, characterized by a strong directional component and partial ionic polarization. Such bonding combines strong directional covalency with enhanced interatomic interaction, thereby significantly improving both electron localization and bonding strength at the GB, effectively suppressing crack initiation and propagation. These results suggest that Si segregation introduces partial covalent bonding character, transforming the originally metallic GB into a hybrid−bonded interface, thus achieving GB strengthening. In the Si+Mg

co−segregation system (Fig. 11(c)), the electron localization distribution exhibits more complex characteristics. The Mg atom, with a larger atomic radius and lower electronegativity than Si, primarily forms metallic bonds upon segregation near the GB, corresponding to relatively weak electron localization. In contrast, the polar covalent bonding between Si and Y leads to locally enhanced electron localization. The synergistic co−segregation of Si and Mg optimizes the electronic structure at the GB: Mg, with its lower electronegativity and larger size, facilitates local strain relaxation and enhances geometric stability through metallic interactions, while Si markedly strengthens electron localization and interfacial cohesion via directional polar covalent bonding with Y. The combined effect of these two solutes results in a composite bonding structure within the GB region, which simultaneously provides ductility and bonding strength, ultimately achieving overall GB stabilization.

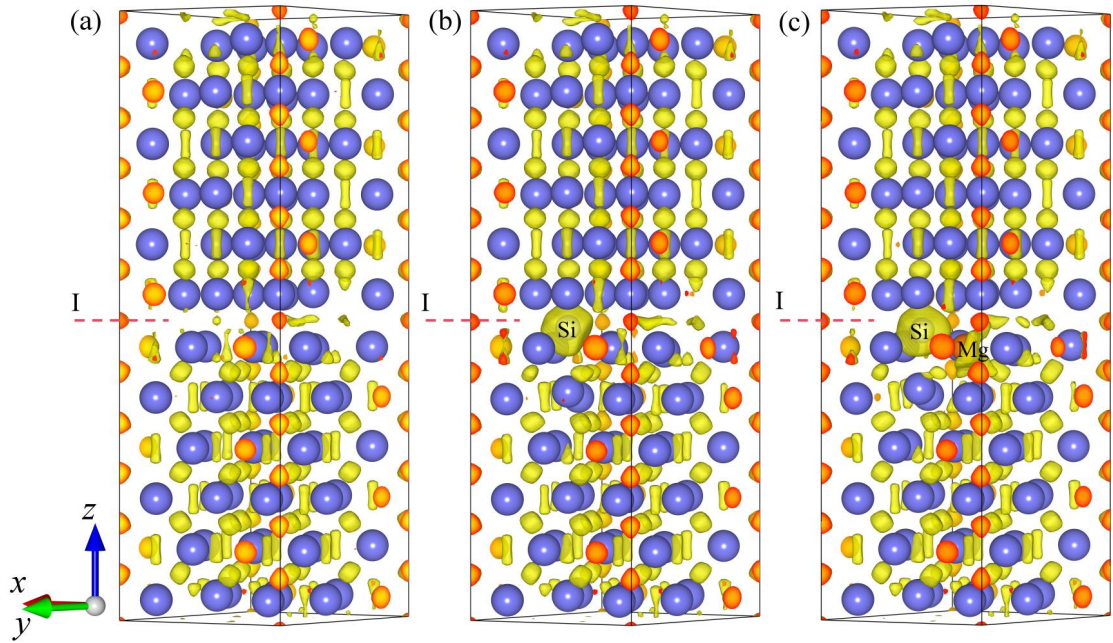

**Figure 11**. Three−dimensional visualization of the electron localization function (isosurface = 0.64) for the pure Yttrium Σ7(0001) twist grain boundary (a), Si−segregated grain boundary (b), and Si–Mg co−segregated grain boundary (c). The dashed line labeled as "I" indicates the preferred fracture path of the grain boundary, corresponding to Path I in Fig. 1.

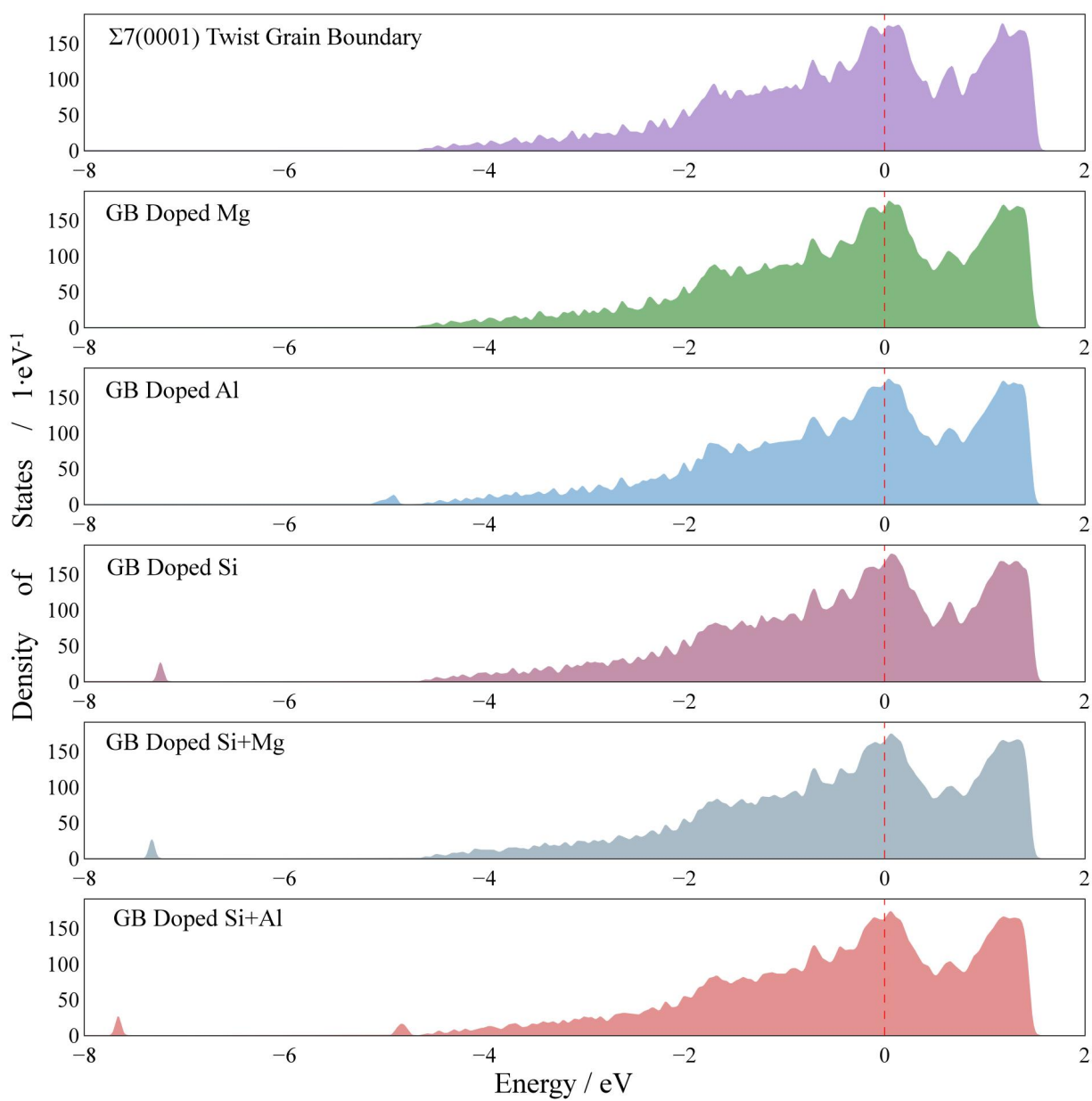

**Figure 12.** DOS of the Σ7(0001) twist grain boundary and the GB systems with partial solute segregation. The segregated systems are labeled in the upper left corner of each subfigure. From top to bottom: the pure Σ7(0001) twist GB, Mg−segregated GB, Al−segregated GB, Si−segregated GB, Si+Mg co−segregated GB, and Si+Al co−segregated GB.

The density of states (DOS) reflects the distribution of electronic states over different energy levels, thereby revealing how solute segregation affects the electronic structure and stability of the grain boundary (GB) [23,58]. Fig. 12 presents the TDOS of the Σ7(0001) twist GB and those of the GB systems with Mg, Al, Si, Si+Mg, and Si+Al segregation. For the pure Σ7(0001) GB system, a pronounced DOS peak appears near the Fermi level ($x = 0$ eV), indicating a high density of electronic states and significant electron delocalization in the GB region, characteristic of metallic bonding. Such weak electronic localization is typically associated with low bonding strength, consistent with the earlier conclusion that the pure GB exhibits weak strength and is prone to fracture. Upon Mg segregation, the DOS peak above the Fermi level ($x \approx 0.66$ eV) decreases noticeably, suggesting that Mg segregation partially optimizes the GB electronic structure and enhances thermodynamic stability (the GB energy decreases by 0.021 J·m$^{-2}$), although its effect remains relatively limited. In the case of Al segregation, a distinct DOS peak emerges in the deep−energy region ($x \approx -4.9$ eV), indicating that Al segregation introduces additional low−energy electronic states. This redistribution of electrons enhances the GB stability (GB energy decreases by 0.101 J·m$^{-2}$), although the bonding remains primarily metallic in nature. The GB system with Si segregation exhibits more pronounced changes in the electronic structure, with a new and distinct DOS peak appearing in the deep-energy region ($x \approx -7.25$ eV). This indicates that Si, owing to its higher electronegativity and smaller atomic radius, can form directional metal–nonmetal hybrid bonds with Y atoms that possess a strong covalent component. Such bonding substantially enhances the GB bonding energy and structural stability, as evidenced by a decrease in GB energy of 0.184 J·m$^{-2}$. In the Si+Mg co−segregation system, the synergistic interaction between Mg and Si causes the DOS peak in the

deep−energy region to shift further toward more negative energy levels ($x \approx -7.33$ eV). This downward energy shift reflects an increase in bonding strength and serves as a direct electronic signature of further GB stabilization. Similarly, in the Si+Al co−segregation system, the DOS peak also shifts toward deeper energy levels ($x \approx -7.67$ eV), indicating a pronounced bond−strengthening effect induced by co−segregation. The cooperative interaction between Al–Y metallic bonds and Si–Y polar covalent bonds further enhances electron coupling and interfacial cohesion within the GB region, thereby achieving maximum grain boundary stabilization.

## 4. Conclusions

In this study, We systematically investigate the segregation and co−segregation behaviors of eleven low−neutron−absorption elements (Si, Cu, Cr, Mo, Zn, Fe, Nb, Mg, Al, Ti and Zr) at the Σ7(0001) twist grain boundary in metallic yttrium, as well as their effects on interfacial stability and mechanical properties. Comparisons of the fracture energy difference between two cleavage paths and first−principles tensile tests between the grain boundary and the bulk reveal that the Σ7(0001) boundary exhibits a certain degree of weakening, with fracture preferentially occurring along path I. Segregation energy analyses show that Si, Cu, Cr, Mo and Fe tend to occupy interstitial sites, whereas the other elements prefer substitutional positions. The overall segregation tendency follows the order: Si > Cu > Cr > Mo > Zn > Fe > Nb > Mg > Al > Ti > Zr. Further results indicate that Si, Al, Zn, Cu, Mg and Fe contribute to enhancing grain boundary stability, while Mo, Fe, Si, Cr, Cu, Nb and Ti markedly strengthen the boundary. Among them, Si demonstrates the most balanced improvement in both aspects. Co−segregation studies reveal that pre−segregated Si induces the enrichment of other solute atoms at the boundary and promotes synergistic stabilization, transforming initially embrittling elements (Al, Mg, Zn, Zr) into boundary−strengthening agents. ELF analyses indicate that Si segregation enhances electron localization at the boundary through the formation of Si–Y covalent bonds, while Si+Mg co−segregation further optimizes the electronic distribution via the cooperative interaction of metallic and covalent bonds, thereby significantly improving fracture strength. The DOS analysis reveals the emergence of new deep−level states in the Al, Si, segregation and Si+Al, Si+Mg co−segregation systems, where enhanced low−energy states contribute to lowering the grain boundary energy and improving structural stability. This study provids theoretical guidance for designing high–performance, low−neutron−absorption Y−based alloys.

**Declaration of competing interest**

The authors declare that they have no known competing financial interests or personal relationships that could have appeared to influence the work reported in this paper.

**Acknowledgments**

This research was funded by the National Key Research and Development Program of China (Grant No. 2023YFB3506704), the National Natural Science

Foundation of China (Grant No. 52371003), and the Fundamental Research Funds for the Central Universities (Grant No. FRF-IDRY-GD24-004).

**Data Availability Satement**

The data that support the findings of this study are available from the corresponding author upon reasonable request.